\documentclass[8pt]{article}

\usepackage{geometry}
\usepackage{array}
\usepackage{ragged2e} 

\usepackage{amsmath, amsthm, amssymb}
\usepackage{graphicx}
\usepackage{hyperref}
\usepackage[utf8]{inputenc}
\usepackage{microtype} 
\usepackage{tabularx}
\usepackage{tikz}
\usetikzlibrary{arrows.meta}
\usetikzlibrary{arrows.meta,decorations.text,calc}
\usepackage[table]{xcolor}
\usetikzlibrary{angles,quotes,patterns,intersections}
\usepackage{xcolor}
\usetikzlibrary{3d}
\usepackage{hhline}
\usepackage{hyperref}

\usepackage{natbib}

\usepackage{times} 
\usepackage{natbib}
\usepackage{setspace}
\theoremstyle{definition} 

\begin{document}

\title{Diagonalization Without Relativization\\\large A Closer Look at the Baker-Gill-Solovay Theorem}
\author{Baruch Garcia\footnote{baruchgarcia@utexas.edu}}
\date{\today}

\maketitle

\begin{abstract}
We already know that several problems like the inequivalence of P and EXP as well as the undecidability of the acceptance problem and halting problem relativize. However, relativization is a limited tool which cannot separate other complexity classes. What has not been proven explicitly is whether the Turing-recognizability of the acceptance problem relativizes. We will consider an oracle for which R and RE are equivalent; $R^A=RE^A$, where A is an oracle for the equivalence problem in the class ALL, but not in RE nor co-RE. We will then differentiate between relativization and what we will call "semi-relativization", i.e., separating classes using only the acceptance problem oracle. We argue the separation of R and RE is a fact that only "semi-relativization" proves. We will then "scale down" to the polynomial analog of R and RE, to evade the Baker-Gill-Solovay barrier using "semi-relativized" diagonalization, noting this subtle distinction between diagonalization and relativization. This "polynomial acceptance problem" is then reducible to CIRCUIT-SAT and 3-CNF-SAT proving that these problems are undecidable in polynomial time yet verifiable in polynomial time. "Semi-relativization" does not employ arithmetization to evade the relativization barrier, and so itself evades the algebrization barrier of Aaronson and Wigderson. Finally, since semi-relativization is a non-constructive technique, the natural proofs barrier of Razborov and Rudich is evaded. Thus the separation of R and RE as well as P and NP both do not relativize but do "semi-relativize", evading all three barriers. 
\end{abstract}

\tableofcontents
\newpage
\section{Prologue}

\begin{quotation}
    \textit{La verdad adelgaza y no quiebra}...\footnote{"The truth may be stretched thin, but does not break[...]"}- Don Quixote de La Mancha, Miguel de Cervantes
\end{quotation}
To clearly understand the foundations of computational complexity, we will examine the foundations of computability as discussed by Stephen Cook in the early 1970's\cite{Cook_2018_Heidelberg}: 
\begin{quotation}
    My advisor as a graduate student, Hao Wang at Harvard, was interested in logic and recursion theory and what led me to this idea was exactly the theory of computable functions, which had the so-called recursively enumerable sets and so the same phenomenon was there. You had recursively enumerable complete sets and they were all equivalent to the hardest recursively enumerable sets, and of course the difference was these sets were not decidable, there was no algorithm to solve them at all, but that was the analogy. It’s different of course for NP-complete sets. They were computable, but not feasibly computable[....]It presumably takes exponential amount of time to solve the problems, but the analogy is there. That’s where I got the idea.
\end{quotation}

Wigderson\cite{wigderson2019mathematics} further explains the analog:
\begin{quotation}
    A very appealing feature of the $\textbf{P}$ versus $\textbf{NP}$ question (which was a source of early optimism about its possible quick resolution) is that it can be naturally viewed as a \textit{bounded} analog of the decidability question from computability theory[....] To see this, replace the \textit{polynomial-time} bound by a \textit{finite} bound in both classes. For $\textbf{P}$, the analog becomes all problmes having finite algorithms, namely the decidable problems sometimes called \textit{Recursive} problems and denoted by $\textbf{R}$. For $\textbf{NP}$, the analog is the class of properties for which membership can be certified by a finite witness via a finite algorithm. This class is called \textit{Recursively Enumerable}, or $\textbf{RE}$.  
\end{quotation}

We will talk of \textit{finite verification} or equivalently \textit{recursively enumerability} or \textit{computable enumerability} or \textit{semi-decidability} or \textit{Turing-recognizability}. Take the halting problem, for example; although there is no generalized algorithm for the halting problem, if it does halt, it can be verified in finite time $t$. We will refer to this class as: $$\textbf{RE}$$. Its polynomial analog is always referred to as  "non-deterministic polynomial time" or: $$\textbf{NP}$$

We will also use the terms \textit{finite co-verification} equivalently to \textit{co-recursive enumerability} or \textit{co-computable enumerability} or \textit{co-semi-decidability} or \textit{co-Turing-recognizability}. Take the complement of the halting problem, for example; although there is no generalized algorithm for the complement of the halting problem ("does the Turing machine \textit{never} halt?"), its complement is verifiable, i.e. if it halts, it can be verified in finite time $t$. We will refer to this class as: $$co-\textbf{RE}$$ Its polynomial analog of the complement is: $$co-\textbf{NP}$$

We will also talk of \textit{recursion} or equivalently \textit{decidability} or equivalently \textit{computability} or \textit{algorithmic}.  We will refer to this class as: $$\textbf{R}$$ Problems outside this class are said to be \textit{non-recursive} or \textit{undecidable} or \textit{uncomputable}\footnote{"Decidable" may refer to a two-valued decision and "computable" or "solvable" may refer to a multi-valued solution, which can be made of several binary "decisions".} or may be said to have \textit{no generalized algorithm}. Its polynomial analog is referred to as "polynomial time" or $$\textbf{P}$$

This class includes all problems for which there is a generalized algorithm to decide (or solve) in a \textit{feasible}, i.e., polynomial, amount of time. In this paper, we will rely heavily on notation used in Sipser's \textit{Introduction to the Theory of Computation} (3rd edition), differing notably in our use of $\textbf{RE}$ for "Turing-recognizability" and $\textbf{R}$ for "decidability".

\section{Background}
Gödel's letter to von Neumann\cite{Godel1956Letter}\footnote{John Nash discovered a partial version of the problem involving the infeasibility of $\textbf{NP}-intermediate$ problems\cite{aaronson2016p}.} established a problem later recognized as the "$\textbf{P}\; \text{versus}\; \textbf{NP}$" problem. Can a the class of problems that can be verified in polynomial, i.e. "feasible", time, also be solved (decided) in polynomial time? Reviews of the problem can be found by Sipser\cite{sipser1992history}, Cook\cite{cook2000p} and Aaronson\cite{aaronson2016p}. Textbooks by Sipser\cite{sipser1996introduction}, Wigderson\cite{wigderson2019mathematics}, Arora and Barak\cite{arora2009computational}, and Aaronson\cite{aaronson2013quantum} also shed light on the problem.

It is important to note that if $\textbf{P}\neq \textbf{NP}$, Ladner's theorem\cite{Ladner1975} tells us there will be an intermediate class, i.e., $\textbf{NP}-intermediate$ which must exist. While it is conceivable that natural problems like graph isomorphism, integer factorization, and discrete logarithm, may change\footnote{...as the problem of primality changed from an $\textbf{NP}-intermediate$ problem to problem in $\textbf{P}$, with the algorithm of Agrawal, Kayal, and Saxena\cite{agrawal2004primes}} into either $\textbf{P}$ or $\textbf{NP}-complete$, the $\textbf{NP}-intermediate$ class would exist in perpetuity, as we see with the $\textbf{RE}-intermediate$ class, which contains no natural problems\cite{Post1944}\cite{KleenePost1954}\cite{Friedberg1957}\cite{Muchnik1956}\cite{aaronson2013quantum}. If $\textbf{P}\neq \textbf{NP}$, then the problems which will always be undecidable in feasible time are called "$\textbf{NP}-complete$". $3-CNF-SAT$---otherwise known as $3SAT$ or $SAT$ for short---is a canonical problem\footnote{An example of of an $\textbf{NP}-complete$ problem is $3-CNF-SAT$ which asks whether the following statement is satisfiable: $$(x_1 \vee \neg x_2 \vee x_3) \wedge (\neg x_1 \vee x_3 \vee \neg x_4) \wedge (x_2 \vee \neg x_3 \vee x_4)$$} from which we provide polynomial-time reductions to prove that other well-known problems, like $CLIQUE$ and $SUDOKU$, are also $\textbf{NP}-complete$. However, if $\textbf{P}$ were to equal $\textbf{NP}$, then $\textbf{P}$, $\textbf{NP}-intermediate$ and $\textbf{NP}-complete$ problems would collapse to one class. The inequivalence of $\textbf{P}$ and $\textbf{NP}$ seems so obvious for many reasons. For example, there are a multitude of $\textbf{NP}-intermediate$ problems, yet not a single one has been found to have a polynomial-time decision algorithm. A second reason is that, although there are quantum polynomial-time ($\textbf{BQP}$) algorithms for a subset\cite{aaronson2022much} of $\textbf{NP}-intermediate$ problems like Shor's algorithm\cite{shor1999polynomial} for factorization, Bennett, Bernstein, Brassard and Vazirani\cite{bennett1997strengths} famously proved that we cannot have exponential speed-ups for $\textbf{NP}-complete$ problems. In the best possible case, we have a quadratic (polynomial) speed-up, as was famously demonstrated by Grover\cite{grover1997quantum}. So not only, would we expect $\textbf{P}\neq \textbf{NP}$ in classical computers, but quantum computers as well\cite{aaronson2005np}. Thirdly, it goes against "common sense", for whatever that phrase is worth, to assume that every feasibly verifiable problem can also be feasibly solved. It appears the only benefit of even discussing the possibility that $\textbf{P}$ and $\textbf{NP}$ could be equivalent would be to spur a proof of their \textit{inequivalence}.    

Lamentably, three barriers have appeared that make the problem of separating $\textbf{P}$ and $\textbf{NP}$ difficult to tackle with known techniques used to prove other results like $\textbf{P}\neq \textbf{EXP}$, $\textbf{IP} = \textbf{PSPACE}$ and $PARITY\notin \textbf{AC}^0$. Baker, Gill and Solovay(BGS)\cite{baker1975relativizations}, provided an argument to show that some results hold, such as the undecidability of the halting problem or the important result\cite{hartmanis1965computational} that $\textbf{P}\neq \textbf{EXP}$, no matter which oracle one may choose. These results "relativize". And other results do not relativize. For example, one can choose two oracles, which give different results for the relation of $\textbf{P}$ and $\textbf{NP}$. One can choose an oracle $B$ which gives us the anticipated result that $\textbf{P}$ and $\textbf{NP}$ are inequivalent:

$$\textbf{P}^B\neq \textbf{NP}^B$$

However, as BGS show, one can also choose an oracle A which is an $\textbf{EXP}$ (or $\textbf{PSPACE}$) oracle, where the unanticipated equivalence of $\textbf{P}$ and $\textbf{NP}$ can be demonstrated.

$$\textbf{P}^A=\textbf{NP}^A$$

No technique using "relativization" can separate the classes. Later, certain valid results were shown not to relativize. For example, Shamir\cite{shamir1992ip} working on interactive proofs, proved that $IP=PSPACE$, and this result circumvents the relativization barrier, using a process called "arithmetiziation". Arithmetization is the technique of translating Boolean functions into polynomials\footnote{An example of arithmetizing a Boolean statement may be as follows $$(x\land y)\neg x \implies xy-x^2y$$} building upon the following rules\footnote{...as one encounters in probability theory.}:
$$x\land y\equiv x y\;\;\;\;\;\;\;\; x\lor y\equiv x+y-xy\;\;\;\;\;\;\;\;\neg x\equiv1-x$$ 

The algebraic structure is not captured by the more simplistic oracles of Baker, Gill, and Solovay. Alas, Aaronson and Wigderson\cite{aaronson2009algebrization} show that such a technique using arithmetization cannot separate $\textbf{P}$ and $\textbf{NP}$, since an "algebraic oracle" $\widetilde{A}$, capturing the more subtle algebraic structure, can be constructed to show that:

$$\textbf{P}^{\widetilde{A}}=\textbf{NP}^{\widetilde{A}}$$

Finally, one may try to abandon the attempt to search "globally" and instead search "locally", restricting arguments to the bit-level. More specifically, one may give up non-constructive proofs like diagonalization arguments for constructive proofs which rely on circuit complexity arguments that have been successful to prove results like $PARITY\notin \textbf{AC}^0$\cite{furst1984parity}. However, Razborov and Rudich showed one runs into a "self-defeating" \textbf{third} barrier called the "natural proofs" barrier going this route. Aaronson explains\cite{aaronson2016p}:

\begin{quotation}
    As Razborov and Rudich themselves stressed, the take-home message is not that we should give up on proving $\textbf{P}\neq\textbf{NP}$. In fact, since the beginning of complexity theory, we’ve had at least one technique that easily evades the natural proofs barrier: namely, diagonalization [....] Of course, diagonalization is subject to the relativization barrier [....] so the question still stands of how to evade relativization and natural proofs simultaneously [....]
\end{quotation}

What is needed is a proof that evades all three barriers like the result of Williams that $\textbf{NEXP} \not\subset \textbf{ACC}^0$\cite{Williams2014}. The program of "meta-complexity" also aims to evade the natural proofs barrier\cite{ilango2022robustness}. 

We see in the problem of $\textbf{P}$ versus $\textbf{NP}$ and even its barriers, a common thread of self-reference, which Ketan Mulmuley's program of Geometric Complexity Theory (GCT)\cite{mulmuley2001geometric} has tried to break. However, this program has been proposed to take 100 years\cite{fortnow2009status}, so we will take time to revisit, in greater depth. the simplest barrier of relavitization as established by Baker, Gill and Solovay\cite{baker1975relativizations}.

\section{Back to Basics from Complexity to Computability}

Let us define an algorithm and Turing machine. Wigderson\cite{wigderson2019mathematics} explains:
\begin{quotation}
    \textit{Turing defined an \textbf{algorithm} (also called a} decision procedure\textit{) to be a Turing machine (in modern language, simply a computer program) that halts on every input in finite time}. 
\end{quotation}

A \textbf{Turing machine} is formally defined\cite{sipser1996introduction} as a 7-tuple: $$(Q,\Sigma,\Gamma,\delta,q_0,q_{acc},q_{rej})$$

\begin{itemize}
    \item $Q$ is the set of states.
    \item $\Sigma$ is the input alphabet, not containing the blank symbol $\textvisiblespace$ .
    \item $\Gamma$ is the tape alphabet, containing the blank symbol $\textvisiblespace$ . 
    \item $\delta:Q \times \Gamma \times\{L,R\}$ is the transition function, moving the head left $L$ or right $R$. 
    \item $q_0 \in Q$ is the start state.
    \item $q_{acc} \in Q$ is the state where the Turing machine \textbf{halts} and \textbf{accepts}.
    \item $q_{rej} \in Q$ is the state where the Turing machine \textbf{halts} and \textbf{rejects}.
\end{itemize}

   Historically, the undecidability of the halting problem discovered by Turing\cite{turing1936computable}---on the heels of Gödel's demonstration of the existence of undecidable propositions in Peano arithmetic\cite{Goedel1931}---reigns supreme as the foundation and gold standard for computability theory. However, Sipser\cite{sipser1996introduction} has correctly identified the \textit{acceptance} problem, which he denotes as $A_{TM}$ as the more general case, from which one can reduce the proof of the undecidability of the halting problem. The acceptance problem is defined as follows:

$$A_{TM}=\{\langle M, w \rangle|\: M\;\text{is a}\;TM\;\text{and}\; M\;\text{accepts}\;w\}.$$
   
   See, the halting problem presumes that a Turing machine even \textit{has} the option to never halt, contrary to its design \textit{to} halt. Turing knew from Gödel's 1931 discovery that not all Turing machines \textit{would} halt. But instead of beginning with Gödel's discovery, one can begin with a proof for the undecidability of the acceptance problem, as Sipser does.

\begin{figure}[htbp]
    \centering

\centering

\begin{center}
\resizebox{0.7\textwidth}{!}{
\begin{tabular}{|*{6}{c|}} 
\hline
  & $\ulcorner M_{1} \urcorner$ & $\ulcorner M_{2} \urcorner$ & $\ulcorner M_{3} \urcorner$ & $\ldots$ & $\ulcorner D \urcorner$ \\ \hline
$M_{1}$ & \cellcolor{lightgray}Accept & Accept & Reject & $\ldots$ & Reject \\ \hline
$M_{2}$ & Reject & \cellcolor{lightgray}Reject & Accept & $\ldots$ & Accept \\ \hline
$M_{3}$ & Accept & Reject & \cellcolor{lightgray}Reject & $\ldots$ & Reject \\ \hline
$\vdots$ & $\vdots$ & $\vdots$ & $\vdots$ & \cellcolor{lightgray}$\ddots$ & $\ldots$ \\ \hline
$D$ & \cellcolor{lightgray}Reject & \cellcolor{lightgray}Accept & \cellcolor{lightgray}Accept & \cellcolor{lightgray}$\ldots$ & \cellcolor{lightgray}? \\ \hline
\end{tabular}
}
\end{center}

    \caption{Diagonalization for the acceptance problem. Now that we know, from the bottom-right cell, that not all Turing machines will halt, we can ask Turing's question of whether or not there exists a generalized algorithm to decide whether a Turing machine halts or \textit{never} halts. As Turing proved, there is no such algorithm. }
    \label{fig:enter-label}
\end{figure}

Now what does the acceptance problem tell us about the separation of the classes $\textbf{R}$ and $\textbf{RE}$?  It tells us that there exists a problem that is not decidable, so not in $\textbf{R}$, \textit{but} the problem can be verified in finite time, putting it in $\textbf{RE}$.  What does it mean to \textit{verify} in finite time? It means that although there is no generalized algorithm to solve the acceptance problem, if it is claimed that a Turing machine accepts, then it must halt in finite time, by definition, so it can be verified\footnote{Note that the complement of the acceptance problem $\overline{A}_{TM}$ is not the same as the recursively enumerable rejection problem $REJECT_{TM}$, because the complement of acceptance is either rejection or never halting (looping).}.

From the undecidability of the acceptance problem, we can establish the complexity class $\textbf{RE}$. We can also consider \textbf{oracle} Turing machines. For example, an oracle machine for the acceptance problem, which Sipser denotes $T^{A_{TM}}$ \textit{can} give a generalized algorithm for the acceptance problem from "outside", avoiding self-reference, by just choosing "\textit{ACCEPT}" or "\textit{REJECT}" for the bottom-right cell in Figure 1. However, when applying the oracle to itself, the same undecidability of the original acceptance problem is encountered. The acceptance problem is---in this way---always outside the class $\textbf{R}$, not matter which oracle is chosen---we say that the undecidability of the acceptance problem \textit{relativizes}\footnote{... as does the undecidability of the halting problem, using the same argument.}.

\section{Relativization Versus "Semi-Relativization"}

We know from Baker, Gill, and Solovay\cite{baker1975relativizations}\cite{moshkovitz2012lecture} that we can make an $\textbf{EXP}$ (exponential time\footnote{Note that just as $\textbf{R}\neq \textbf{ALL}$ relativizes so does $\textbf{P}\neq \textbf{EXP}$ relativize.}) oracle $A$ such that:

$$\textbf{P}^{A}=\textbf{NP}^{A}$$

So we know that any proof of $\textbf{P}\neq \textbf{NP}$ will not relativize. Let us now look at the analog of oracle $A$ for $\textbf{R}$ and $\textbf{RE}$. Let us first establish that $\textbf{R}$ and $\textbf{RE}$ are indeed separate. We know from the undecidability of the acceptance problem and the halting problem that there exists a problem that is in $\textbf{RE}$, but not in $\textbf{R}$, separating the two classes. The complement of the acceptance problem, which Sipser calls $\overline{A}_{TM}$, is in the complement of $\textbf{RE}$, i.e. "$co-\textbf{RE}$". Now there is a class which is a strict superset of the aforementioned languages and indeed, all languages, called "$\textbf{ALL}$", otherwise known as "$\textbf{LANG}$". $\textbf{ALL}$ includes the "equivalence problem" that Sipser\cite{sipser1996introduction} calls $EQ_{TM}$ and places in this class that has an uncountably infinite number of languages. The equivalence problem $EQ_{TM}$ is defined as follows:

$$EQ_{TM}=\{\langle M_1, M_2 \rangle|\: M_1\; \text{and}\; M_2\;\text{are TMs and}\;L(M_1)=L(M_2)\}.$$

Sipser\cite{sipser1996introduction} shows in Corollary 5.29 that one can prove $B$ is not Turing-recognizable (or "recursively enumerable" $\textbf{RE}$) if one can show that the acceptance problem reduces to the complement of problem $B$:

$$A_{TM}\leq\overline{B}$$

Sipser then demonstrates the both $EQ_{TM}$ and $\overline{EQ}_{TM}$ can be reduced from the acceptance problem $A_{TM}$ with simple strategy, proving that $EQ_{TM}$ is neither Turing-recognizable nor co-Turing-recognizable, i.e., $EQ_{TM}$ is neither in $\textbf{RE}$ nor $co-\textbf{RE}$. To show that $EQ_{TM} \notin \textbf{RE} $, Sipser shows that $A_{TM}\leq\overline{EQ}_{TM}$ by the following argument:
\begin{itemize}
    \item Choose $M_1$ that rejects on any input.
    \item Choose $M_2$ that runs $M$ on $w$, on any given input. If it accepts, $accept$. 
\end{itemize}
If $M$ rejects $w$, then it always rejects, and $M_1$ and $M_2$ are equal. If $M$ accepts $w$, then it always accepts, and $M_1$ and $M_2$ are not equal. Then to show that $EQ_{TM} \notin co-RE$, just prove that $A_{TM}\leq{EQ}_{TM}$ by the following argument:
\begin{itemize}
    \item Choose $M_1$ that accepts on any input.
    \item Choose $M_2$ that runs $M$ on $w$, on any given input. If it accepts, $accept$. 
\end{itemize}

From the equivalence problem, we can create an \textbf{equivalence problem oracle} in an analogous way we would do for an acceptance problem oracle or a halting problem oracle. While the equivalence problem oracle can tell us the answer to the equivalence problem, there is no generalized algorithm that answers the equivalence problem for an oracle machine equivalent to itself. We now have an oracle that can solve the acceptance problem, but is neither Turing recognizable nor Turing co-recognizable: 

$$T^{EQ_{TM}}\notin \textbf{RE}\;\;\;\;\;\text{and}\;\;\;\;\;T^{EQ_{TM}}\notin co-\textbf{RE}$$

In summary we have the following problems and oracles: 
\begin{itemize}
    \item \textbf{Acceptance Problem} ($A_{TM}$) is $\textbf{RE}-complete$: Undecidable for a Turing machine, but verifiable in finite time if Turing machine \textit{does} accept.
    \item \textbf{Acceptance Oracle} ($T^{A_{TM}}$) is $\textbf{RE}-complete$: Undecidable for equivalent oracle machine, but verifiable in finite time if the oracle \textit{does} accept.
    \item \textbf{Equivalence Problem} ($EQ_{TM}$) is in $\textbf{ALL}$: Undecidable; not verifiable, nor co-verifiable. 
    \item \textbf{Equivalence Oracle} ($T^{EQ_{TM}}$) in $\textbf{ALL}$: Can decide the accpetance problem, but any equivalent oracle is undecidable; not verifiable, nor co-verifiable in finite time. 
\end{itemize}

We \textit{can} show the anticipated result:

$$\textbf{R}^B\neq \textbf{RE}^B$$

where $B$ is an oracle in $\textbf{RE}$, such as the acceptance oracle, so:

$$\textbf{R}^{A_{TM}}\neq \textbf{RE}^{A_{TM}}$$

Which is not surprising and to be expected. But then we can have the more surprising result highlighted in this paper, that: 

$$\textbf{R}^A=\textbf{RE}^A=co-\textbf{RE}^A$$

where $A$ is an oracle in $\textbf{ALL}$, and we can use specifically the equivalence oracle, so:

$$\textbf{R}^{EQ_{TM}}=\textbf{RE}^{EQ_{TM}}=co-\textbf{RE}^{EQ_{TM}}$$

because the acceptance problem is now decidable, but the equivalence oracle itself is undecidable, but in $\textbf{ALL}$. If the equivalence oracle decides the acceptance problem or its complement, there is no way to verify nor co-verify the the state of the oracle itself. So we have: So the universally accepted\footnote{Sipser states that $\textbf{R}\subsetneq \textbf{RE}$ in plain language in the text accompanying Theorem 4.11 of \textit{Introduction...}\cite{sipser1996introduction},"[...]recognizers \textit{are} more powerful the deciders." Italics in the original.} statement that:  

$$\textbf{R}\subsetneq \textbf{RE}$$

can only be proven with the acceptance oracle, and not relativization in general, which would include all oracles, include an $\textbf{ALL}$ oracle, like the equivalence oracle. Separating with just the acceptance problem, is what we will call "semi-relativization". So the fact that $R\subsetneq RE$ does not relativize, but does \textit{semi-relativize}, demonstrates the validity of the semi-relativization technique, i.e. diagonalization with the acceptance oracle alone.

\section{"Scaling down" to polynomial-time problems}

We will now define a problem called the "polynomial-time acceptance problem", which asks whether a Turing machine (TM) accepts in polynomial time and is formally defined as:

$$A_{TM-PTIME}=\{\langle M, w \rangle|\: M\;\text{is a}\;TM\;\text{and}\; M\;\text{accepts}\;w\;\text{in}\; \mathcal{O}(n^k)\;\text{time}\}.$$

We can create a diagonalization proof as in Figure 1, but where we have the choices of accepting in \textit{polynomial} time or rejecting in \textit{polynomial} time. Each "$ACCEPT$" cell in Figure 1 would be replaced with "$ACCEPT\;in\;\mathcal{O}(n^k)\; time.$", and each "$REJECT$" cell would be replaced with "$REJECT\;in\;\mathcal{O}(n^k)\; time.$". The bottom-right cell of the diagonalization argument tells the Turing machine to accept in polynomial time if and only if rejects in polynomial time. While there is no generalized algorithm to halt and accept in polynomial time, the Turing machine can \textbf{can} halt, \textit{accepting} or \textit{rejecting}, in $\textbf{EXP}$, i.e.,  $2^{n^k}$ time\footnote{...or longer.}, which is the same as an exhaustive search. However, once an $ACCEPT$ state in the Turing machine, i.e., $q_{acc}$, is reached, it can be verified polynomially, placing the problem outside of $\textbf{P}$ but inside $\textbf{NP}$. We can also define an oracle:
$$T^{A_{TM-PTIME}}$$
Note that this while this result \textit{semi-relativizes}, with a "polynomial-time acceptance oracle", $T^{A_{TM-PTIME}}$, as a canonical example, it does \textit{not} relativize.

 Semi-relativization does not require the methods of arithmetization to evade the relativization barrier. Since semi-relativization evades the relativization barrier without arithmetization techniques, it does not come against the algebrization barrier as techniques for other proofs, e.g., $\textbf{IP}=\textbf{PSPACE}$, do\footnote{Consider the following generalized version of familiar interactive proofs, where we have a powerful but crooked prover $\widetilde{P}$ and a finite but skeptical verifier $V$, following the example of the well-known graph non-isomorphism verification protocol\cite{sipser1996introduction}. Generalizing for finitely verifiable problems may provide supplementary insight into the limitations of interactive proofs. Boone, Novikov and Britton\cite{rotman2012introduction}\cite{turing1950word} proved the following:
\begin{center}
    "There exists a finitely presented group $B$ with an unsolvable (undecidable) word problem."
\end{center}

We will look at the group non-isomorphism problem for finitely presented groups. For an exposition on this problem, see Rotman\cite{rotman2012introduction}. We use rules known as a Thue system\cite{thue1914probleme} (or local canonical Post systems\cite{post1943formal}). 

Particular selections of initial list for which there is no algorithm for deciding do exist. A specific list from Tseitin\cite{tseitin1958associative}\cite{nybergbrodda2024tseytin} is as follows: 

\begin{center}
   (1) AC=CA (2) AD=DA (3) BC=CB (4) BD=DB (5) ECA=CE (6) EDB=DE (7) CCA=CCAE
\end{center}
 
Some relations like:

\begin{center}
    ACCA=ACAC
\end{center}

are easily demonstrated using (1) once, but others are not so obvious. If $\widetilde{P}$  claims two words are isomorphic, then it is easy to show $V$  in a finite number of steps the verification and which rules are used. However, if $\widetilde{P}$  claims the two words are isomorphic, and cannot show $V$  within a finite number of steps the verification, then $V$  is not convinced. Alternately, if $\widetilde{P}$  claims the words are not isomorphic, it seems, at first, impossible for $V$ to verify this claim in a finite time, since the problem reduces from $\overline{HALT}_{TM}$ and $\overline{A}_{TM}$. So, $V$  can choose an interactive proof, where $V$  can flip a coin and secretly choose one word or the other word, and change it according the rules provided. Then $V$  asks $\widetilde{P}$  which word was initially chosen. If the words are isomorphic, then there is a very small chance---$Prob=2^{-N}$, where $N$ is the number of coin flips---for $\widetilde{P}$ to guess correctly every time. If $\widetilde{P}$ answers correctly each time, $V$ can verify that the two words are \textit{not} isomorphic with probability asymptotically approaching unity:

$$Prob=   1-(\frac{1}{2})^N$$ 
}. All relativization is also algebrization. All relativization is also semi-relativization. Crucially, there exist instances of semi-relativization that are neither relativization nor algebrization.

How do we connect $A_{TM-PTIME}$ to canonical $\textbf{NP}-complete$ problems? Let's look at the following "Cook-Levin Tableau" which reduces from $A_{TM-PTIME}$ to a $CIRCUIT-SAT$ problem, which is $\textbf{NP}-complete$, itself reducible to $3-CNF-SAT$\footnote{The Tseitin\cite{tseitin1983complexity} transformation is a simple reduction from $CIRCUIT-SAT$ to $3-CNF-SAT$. Simply build the circuit with $NAND$ gates, with $v_1\; \text{and}\; v_2$ as the inputs and $v_3$ as the output. This will then correspond to the $CNF$ clauses $(v_1\lor v_3)\land(v_2 \lor v_3)\land(\neg v_1 \lor \neg v_2 \lor \neg v_3)$.} 

$$A_{TM-PTIME}\leq  CIRCUIT-SAT\leq 3-CNF-SAT$$

So $A_{TM-PTIME}$, i.e. the problem of whether a Turing machine $M$ accepts some input $w$ in polynomial time, is \textit{not} decidable in polynomial time, but verifiable in polynomial time, since any polynomial-time verification can be represented by a single instance in the Cook-Levin tableau, that ends with $q_{acc}$.

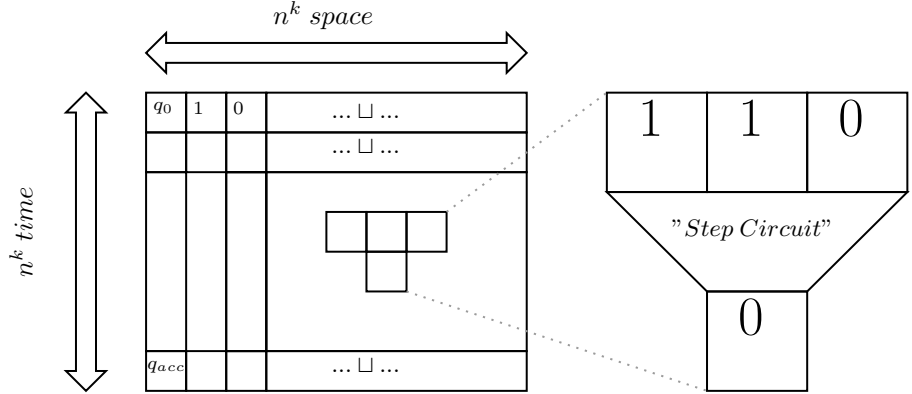
\begin{figure}[htbp]
    \centering

\tikzset{every picture/.style={line width=0.75pt}} 

\begin{tikzpicture}[x=0.75pt,y=0.75pt,yscale=-1,xscale=1]

\draw   (120,70) -- (140,70) -- (140,90) -- (120,90) -- cycle ;
\draw   (140,90) -- (160,90) -- (160,110) -- (140,110) -- cycle ;
\draw   (120,90) -- (140,90) -- (140,110) -- (120,110) -- cycle ;
\draw   (140,70) -- (160,70) -- (160,90) -- (140,90) -- cycle ;
\draw   (160,90) -- (180,90) -- (180,110) -- (160,110) -- cycle ;
\draw   (160,70) -- (180,70) -- (180,90) -- (160,90) -- cycle ;
\draw   (180,70) -- (310,70) -- (310,90) -- (180,90) -- cycle ;
\draw   (180,90) -- (310,90) -- (310,110) -- (180,110) -- cycle ;
\draw   (120,110) -- (140,110) -- (140,200) -- (120,200) -- cycle ;
\draw   (160,110) -- (180,110) -- (180,200) -- (160,200) -- cycle ;
\draw   (160,110) -- (180,110) -- (180,200) -- (160,200) -- cycle ;
\draw   (120,200) -- (140,200) -- (140,220) -- (120,220) -- cycle ;
\draw   (140,200) -- (160,200) -- (160,220) -- (140,220) -- cycle ;
\draw   (160,200) -- (180,200) -- (180,220) -- (160,220) -- cycle ;
\draw   (180,200) -- (310,200) -- (310,220) -- (180,220) -- cycle ;
\draw   (180,110) -- (310,110) -- (310,200) -- (180,200) -- cycle ;
\draw   (210,130) -- (230,130) -- (230,150) -- (210,150) -- cycle ;
\draw   (230,130) -- (250,130) -- (250,150) -- (230,150) -- cycle ;
\draw   (250,130) -- (270,130) -- (270,150) -- (250,150) -- cycle ;
\draw   (230,150) -- (250,150) -- (250,170) -- (230,170) -- cycle ;
\draw   (350,70) -- (400,70) -- (400,120) -- (350,120) -- cycle ;
\draw   (400,70) -- (450,70) -- (450,120) -- (400,120) -- cycle ;
\draw   (400,170) -- (450,170) -- (450,220) -- (400,220) -- cycle ;
\draw   (450,70) -- (500,70) -- (500,120) -- (450,120) -- cycle ;
\draw   (500,120) -- (450,170) -- (400,170) -- (350,120) -- cycle ;
\draw   (120,50) -- (130,40) -- (130,45) -- (300,45) -- (300,40) -- (310,50) -- (300,60) -- (300,55) -- (130,55) -- (130,60) -- cycle ;
\draw   (90,70) -- (100,80) -- (95,80) -- (95,210) -- (100,210) -- (90,220) -- (80,210) -- (85,210) -- (85,80) -- (80,80) -- cycle ;
\draw [color={rgb, 255:red, 155; green, 155; blue, 155 }  ,draw opacity=1 ] [dash pattern={on 0.84pt off 2.51pt}]  (270,130) -- (350,70) ;
\draw [color={rgb, 255:red, 155; green, 155; blue, 155 }  ,draw opacity=1 ] [dash pattern={on 0.84pt off 2.51pt}]  (250,170) -- (400,220) ;

\draw (381,132.4) node [anchor=north west][inner sep=0.75pt]  [font=\small]  {$"Step\ Circuit"$};
\draw (122,73.4) node [anchor=north west][inner sep=0.75pt]  [font=\scriptsize]  {$q_{0}$};
\draw (142,73.4) node [anchor=north west][inner sep=0.75pt]  [font=\scriptsize]  {$1$};
\draw (162,73.4) node [anchor=north west][inner sep=0.75pt]  [font=\scriptsize]  {$0$};
\draw (182,22.4) node [anchor=north west][inner sep=0.75pt]    {$n^{k} \ space$};
\draw (50.4,164) node [anchor=north west][inner sep=0.75pt]  [rotate=-270]  {$n^{k} \ time$};
\draw (211,202.4) node [anchor=north west][inner sep=0.75pt]  [font=\small]  {$...\sqcup ...$};
\draw (119,203.4) node [anchor=north west][inner sep=0.75pt]  [font=\scriptsize]  {$q_{acc}$};
\draw (364,72.4) node [anchor=north west][inner sep=0.75pt]  [font=\huge]  {$1$};
\draw (414,72.4) node [anchor=north west][inner sep=0.75pt]  [font=\huge]  {$1$};
\draw (464,72.4) node [anchor=north west][inner sep=0.75pt]  [font=\huge]  {$0$};
\draw (414,172.4) node [anchor=north west][inner sep=0.75pt]  [font=\huge]  {$0$};
\draw (211,92.4) node [anchor=north west][inner sep=0.75pt]  [font=\small]  {$...\sqcup ...$};
\draw (211,74.4) node [anchor=north west][inner sep=0.75pt]  [font=\small]  {$...\sqcup ...$};

\end{tikzpicture}

    \caption{ A \textit{Cook-Levin Tableau} representing the internal structure of a Turing machine running in polynomial time, with the 7-tuple represented. The "step circuit" has a constant number of logical gates---$AND$, $OR$, $NOT$---and is how we describe the transition function $\delta:Q \times \Gamma \times\{L,R\}$ of the Turing machine as a circuit, moving left $L$ or right $R$ or neither direction on the tape. Note that the top entries of a step circuit can include blanks, and not just 0's and 1's, so a tape can move out to the right and back to the left. Polynomial time and polynomial space with a constant-sized step-circuit gives us a polynomial number of logical gates. $CIRCUIT-SAT$ is not to be confused with with $CIRCUIT-VALUE$ problem which takes the input and in polynomial time \textit{accepts} or \textit{rejects}, placing $CIRCUIT-VALUE$ in $\textbf{P}$, which is the polynomial analog of $\textbf{R}$. $A_{TM-PTIME}$ is reducible to $CIRCUIT-SAT$ through the Cook-Levin Tableau, which itself is reducible to $3-CNF-SAT$ through Tseitin transformations. } 
    \label{fig:enter-label}
\end{figure}

\newpage
\section{Discussion}

We have demonstrated using the methods of "semi-relativized" diagonalization---the simplest form of diagonalization one can imagine---that $\textbf{P}\subsetneq \textbf{NP}$ even though the separation does not relativize; while all relativization is also semi-relativization, not all semi-relativization is relativization. The relativization barrier is evaded. Also, we see that arithmetization---as helpful as it is for other proofs---was not used, so the algebrization barrier is evaded. And since this proof rests on diagonalization, it evades the natural proofs barrier.  
\newpage
\section{Acknowledgements}
The author expresses his gratitude to Scott Aaronson for emphasizing the importance of mathematical rigor, to Michael Sipser for helping clarify early conceptual hurdles, and to Noson Yanofsky for his encouragement to pursue fundamental questions. I am grateful to the BEYOND Center at Arizona State University, and its director, Paul C.W. Davies, for an invitation to visit and present a seminar. The stimulating environment and cross-disciplinary conversations during that visit inspired the development of techniques in this paper. No external funding was received for this work.

\newpage

\bibliographystyle{plain}

\bibliography{references}

\end{document}